\documentclass[aps,prl,twocolumn,groupedaddress,preprintnumbers]{revtex4-1}
\usepackage{amssymb}
\usepackage{graphicx}
\usepackage{subfig}
\usepackage{url}

\def\ie{{\it i.e.}}
\def\eg{{\it e.g.}}

\def\to{\rightarrow}

\def\lsim{\mathrel{\mathpalette\atversim<}}

\newskip\zatskip \zatskip=0pt plus0pt minus0pt
\def\matth{\mathsurround=0pt}
\def\lsim{\mathrel{\mathpalette\atversim<}}

\def\atversim#1#2{\lower0.7ex\vbox{\baselineskip\zatskip\lineskip\zatskip
  \lineskiplimit 0pt\ialign{$\matth#1\hfil##\hfil$\crcr#2\crcr\sim\crcr}}}

\begin{document}

\preprint{SLAC-PUB-16499}

\title{750 GeV Diphoton Resonance in Warped Geometries}


\author{JoAnne L. Hewett and Thomas G. Rizzo}
\email[]{hewett,rizzo@slac.stanford.edu}
\affiliation{SLAC National Accelerator Laboratory, Menlo Park, CA  USA}



\begin{abstract}
We examine the scenario of a warped extra dimension containing bulk SM fields in light of the observed diphoton excess 
at 750 GeV.  We demonstrate that a spin-2 graviton whose action contains localized kinetic brane terms for both gravity 
and gauge fields is compatible with the excess, while being consistent with all other constraints.  The graviton sector 
of this model contains a single free parameter, once the mass of the graviton is fixed. The scale of physics on the 
IR-brane is found to lie in the range of a $\sim$ few TeV,  relevant to the gauge hierarchy. There remains significant 
flexibility in the coupled gauge/fermion KK sectors to address the strong constraints arising from precision measurements.  

\end{abstract}


\maketitle


The large disparity between the electroweak and 4-dimensional scale of gravity has served as inspiration for numerous theories of 
physics beyond the Standard Model (SM), as well as for corresponding experimental searches, for several decades.  One appealing
possibility is that this hierarchy is generated by the geometry of additional spacial dimensions\cite{Hewett:2002hv}, with a 
promising scenario based on gravity being localized in a warped extra dimension\cite{Randall:1999ee}.
The principle signature\cite{Davoudiasl:1999jd} of the warped geometry scheme is the production of Kaluza-Klein (KK) resonances at
high-energy colliders.

The ATLAS and CMS collaborations have recently reported an excess in the diphoton invariant mass spectrum at 750 GeV in both of
the 8 and 13 TeV data sets\cite{LHC750,lhcmoriond}.  While the global significances of these excesses are not (yet) large, being of 
order $2\sigma$ for each data set, 
the presence of an excess in each of these separate collections of data is striking.  The observed cross section for this excess 
differs between the two experiments, with the ATLAS 13 TeV data
preferring larger cross sections of order $\sim 9-10$ fb, while the CMS 13 TeV data 
favors lower values of $\sim 3-5$ fb.  The 8 TeV data sample for both collaborations is found to be 
consistent with the CMS 13 TeV results.  
A useful statistical discussion\cite{Buckley:2016mbr} of the various analyses indicates a slight preference for the lower cross section. 
The possibility that this is the new physics holy grail we have been seeking has sent the theoretical
community into a frenzy\cite{theory750}.  Few of these works\cite{Giddings:2016sfr} have seriously considered the prospect of a 
spin-2 resonance as the source of the excess, and all of these efforts suggest some tension with data.  In this paper, we demonstrate 
that a realistic scenario can be constructed with a warped extra dimension that is naturally consistent with the observation of a 
diphoton resonance at 750 GeV, while satisfying constraints from other data.

This theoretical framework is based on a slice of AdS$_5$ spacetime bounded by two 4-dimensional Minkowski branes with  
the 5$^{th}$ dimension, denoted by $y=r_c \phi$ being bounded by the UV- or Planck(IR- or SM) brane sitting at $y=0(\pi r_c)$. 
The associated 5-D metric is 
\begin{equation}
ds^2 = e^{-2ky}\eta_{\mu\nu}dx^\mu dx^\nu - dy^2 \,.
\end{equation}
The warp factor is represented by the exponential and the parameter $k$, of order the Planck scale, governs the degree of curvature of 
the AdS$_5$ space.  The relation ${\overline M}^2_{Pl}=M^3_5/k$ is derived from the 5-D action and indicates that there are no 
additional hierarchies in this scenario.  
The scale of physical phenomena realized by a 4-D flat metric transverse to the 5$^{th}$ dimension is specified by the 
warp factor, and is given on the IR-brane by $\Lambda_\pi={\overline M}_{Pl}\epsilon$ with $\epsilon= e^{-kr_c\pi}$. Assuming 
$\Lambda_\pi\simeq$ a few TeV to characterize
the electroweak scale on the IR-brane sets $kr_c \simeq 11-12$. (Here we take this to be 11.27.) It has been 
demonstrated\cite{goldbergerwise} that this configuration can be stabilized without the fine tuning of parameters.  The gauge 
hierarchy is thus naturally established by the warp factor $\epsilon \sim 10^{-(15-16)}$.

The simplest picture places all the SM fields on the IR-brane, with only gravity propagating in the extra dimension, and is 
governed by only two parameters $\Lambda_\pi$ and the ratio $k/{\overline M}_{Pl}$.  After compactification, the mass of the $n^{th}$ 
KK graviton state is $m^G_n=x^G_nk\epsilon=x^G_n k\Lambda_\pi/{\overline M}_{Pl}$, with $x^G_n$ being the roots of the 
Bessel function $J_1$. The 
coupling strength for each graviton KK excitation is $\Lambda^{-1}_\pi$\cite{Davoudiasl:1999jd}. In this scenario, 
the graviton interpretation\cite{Giddings:2016sfr} of the diphoton excess has been shown to be in tension with 
the 8 and 13 TeV LHC searches for new states in the dilepton channel. This is because universality of the graviton couplings 
leads to the requirement that $B(G_1\to e^+e^-)=0.5B(G_1\to \gamma\gamma)$ with the 
corresponding scaling of the cross section. Thus with the $\gamma\gamma$ 
production rate being fixed by the observation of the diphoton excess, the predicted dilepton production rate 
lies uncomfortably close to or above the observed 8 and 13 TeV limits\cite{gtension}. 

The additional dimension is small enough that the
SM gauge fields\cite{Davoudiasl:1999tf} and fermions\cite{Davoudiasl:2000wi} are allowed to propagate in the bulk; this scenario 
has attractive model-building features such as the potential to (at least partially) explain the fermion mass hierarchy.  
In this case the masses of the gauge KK states are governed\cite{Davoudiasl:1999tf} by the roots of the equation 
$J_1(x^A_n)+x^A_nJ'_1(x^A_n)+\alpha_n[Y_1(x^A_n)+x^A_nY'(x^A_n)]=0$, with the numerical coefficients $\alpha_n$ being determined  
by the boundary condition on the UV brane. The gauge KK masses are then $m^A_n=x^A_nk\epsilon$.
Comparison of the values of the roots $x^G_n\,, x^A_n$ shows that level by level, the KK excitations
of the gauge bosons are lighter than those of the corresponding graviton KK states.  In particular, the first roots are found to 
take the values $x^G_1 \simeq 3.83$ and $x^A_1 \simeq 2.45$, so that we expect $m^G_1\simeq 1.56 m^A_1$.

When the SM fermions also reside in the bulk, additional parameters ($m_i=k\nu_i$ or $=-kc_i$) are introduced corresponding to the bulk 
fermion masses; 
these parameters determine the fermion wavefunction localization. 
As the Higgs-boson is kept on or near the IR-brane, a consistent picture emerges if light flavor fermions are localized near the 
UV-brane and the third 
generation quarks are localized on or near the IR-brane.  This scenario is severely constrained by precision electroweak
data. However, judiciously enlarging the gauge symmetry to, \eg, the Left-Right Model (LRM), provides\cite{Agashe:2003zs} a custodial 
isospin symmetry in the bulk that 
protects the $\rho$ parameter.  In this scenario constraints from precision measurements, even with the presence of the custodial 
symmetry, still require the first gauge KK states to be heavier than
$\sim 3-4$ TeV\cite{Carena:2007ua}, which in turn pushes the first graviton KK state to be even heavier.

While this scenario paints a nice picture by simultaneously solving the gauge and fermion hierarchies, one more ingredient is needed to 
make contact with the recently observed 750 GeV excess - the introduction of brane localized kinetic terms(BLKTs).  
It has been shown\cite{Georgi:2000ks} that such terms naturally arise from quantum effects.  The phenomenological
consequence of such terms is to modify the spectrum and couplings of the graviton\cite{Davoudiasl:2003zt} and
gauge\cite{Davoudiasl:2002ua} KK states.  

To fix our scenario, we will examine a simplified case where the graviton, gluon, photon and transverse $W$ and $Z$ 
fields propagate in the bulk while the third generation quarks are localized on the IR-brane together with the Higgs and the 
longitudinal components of the $W$ and $Z$, \ie, the Goldstone bosons. 
The graviton and gauge fields have BLKTs on both branes. As shown in the third 
paper in Ref.~\cite{Davoudiasl:2000wi}, once all the light fermions are localized with $\nu=-c \lsim -1/2$ (which we assume) 
they essentially decouple from the KK gravitons, so we have no need to specify their localization further as far as the 
KK gravitons are concerned, however we will return to this issue below. This localization also implies that the $G_1$ decay into 
dilepton pairs will be very highly suppressed.  The 5-D gravitational action is now given by 
\begin{eqnarray}
 S_G = & \frac{M_5^3}{4} & \int d^4 x \int r_c d\phi \, \sqrt {-G} \,
\left\{R^{(5)} + [2\gamma_0/kr_c ~\, \delta(\phi) \right. \nonumber\\ 
& + &
\left. 2\gamma_\pi/kr_c ~\, \delta(\phi - \pi)] \, R^{(4)} + \ldots \right\}\,,
\end{eqnarray}
where $\gamma_{0,\pi}$ represent the kinetic terms for the branes located at $\phi=0,\pi$ and naturally take on values of order 10 when this
model addresses the gauge hierarchy.
The corresponding generic 5-D gauge action can be written as
\begin{eqnarray}
 S_V= & \frac{-1}{4} & \int d^4 x \int r_c d\phi \, \sqrt {-G} \,
\left\{F_{AB}F^{AB} + [2\delta_0/kr_c ~\, \delta(\phi) \right. \nonumber\\ 
& + &
\left. 2\delta_\pi/kr_c ~\, \delta(\phi - \pi)] \, F_{\mu\nu}F^{\mu\nu} + \ldots \right\}\,,
\end{eqnarray}
with $\delta_{0,\pi}$ being the gauge field BLKTs. Here, for simplicity, we will assume that all of the SM gauge fields 
have common values for the kinetic terms although this need not be the case; this has the consequence that KK graviton decays into the 
$Z\gamma$ final state will be absent. We note that the absence of gauge (graviton) ghosts requires 
$\delta_0+\delta_\pi > -\pi kr_c ~(\gamma_0>-1/2)$ while the absence of radion ghosts requires $\gamma_\pi <1$, and we will 
respect these constraints below.   
In the presence of BLKTs, the masses of the graviton KK states are found\cite{Davoudiasl:2003zt} to be governed by the roots of
\begin{equation}
J_1(x^G_n) - \gamma_\pi x^G_n J_2(x^G_n) =0\,,
\end{equation}
with small terms $\sim O(\epsilon^2)$ being dropped.  The KK masses are now functions of $\gamma_\pi$, being
given by $m^G_n=x^G_n(\gamma_\pi)k\epsilon$.  Note that the graviton KK mass spectrum is independent of the BLKT on 
the UV-brane, $\gamma_0$, up to terms suppressed by warp factors. Here, we identify $G_1$ with the 750 GeV excess. 

\begin{figure}[htbp]
\vspace*{-1.0cm}
\includegraphics[scale=0.33,angle=90]{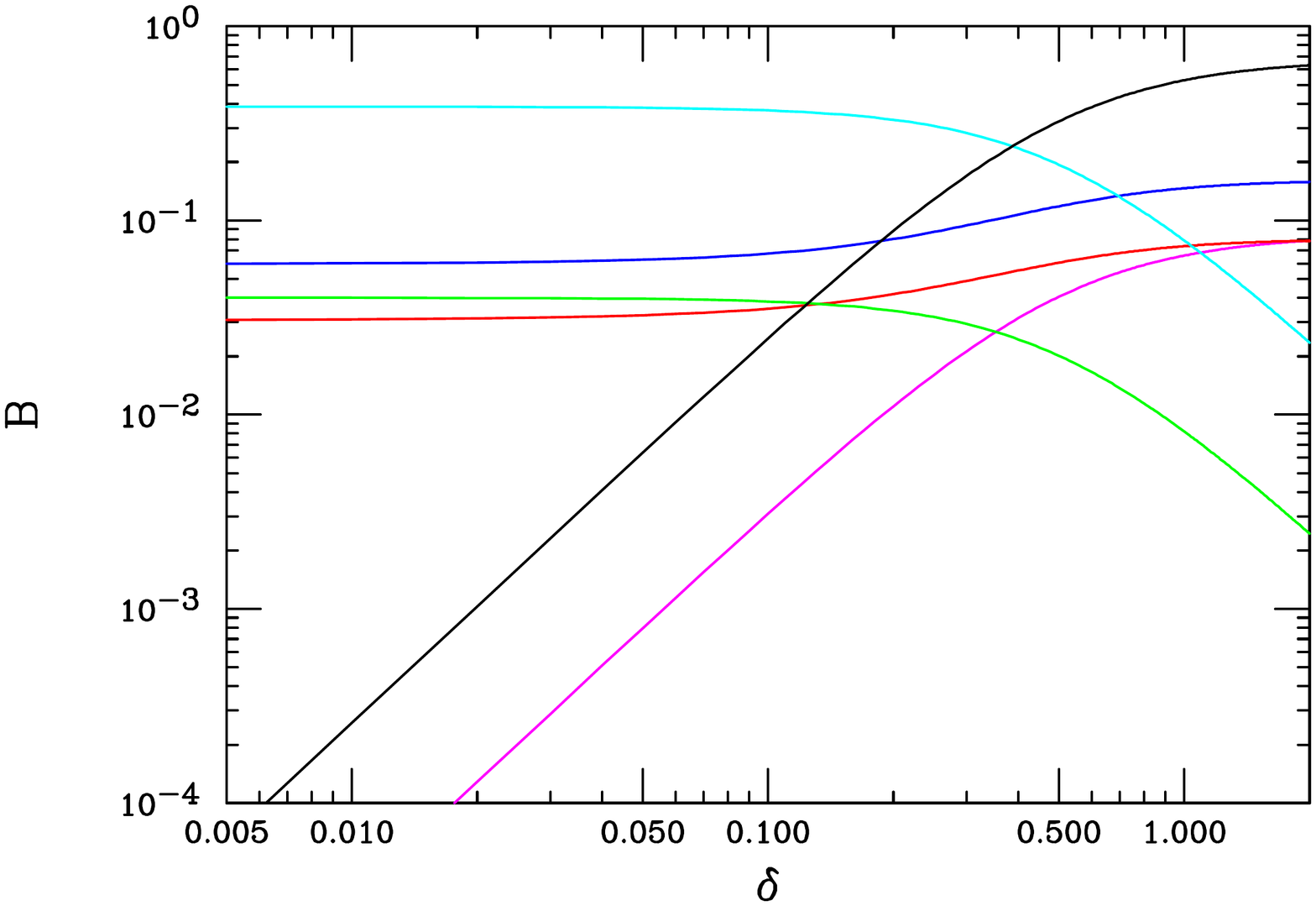}
\includegraphics[scale=0.33,angle=90]{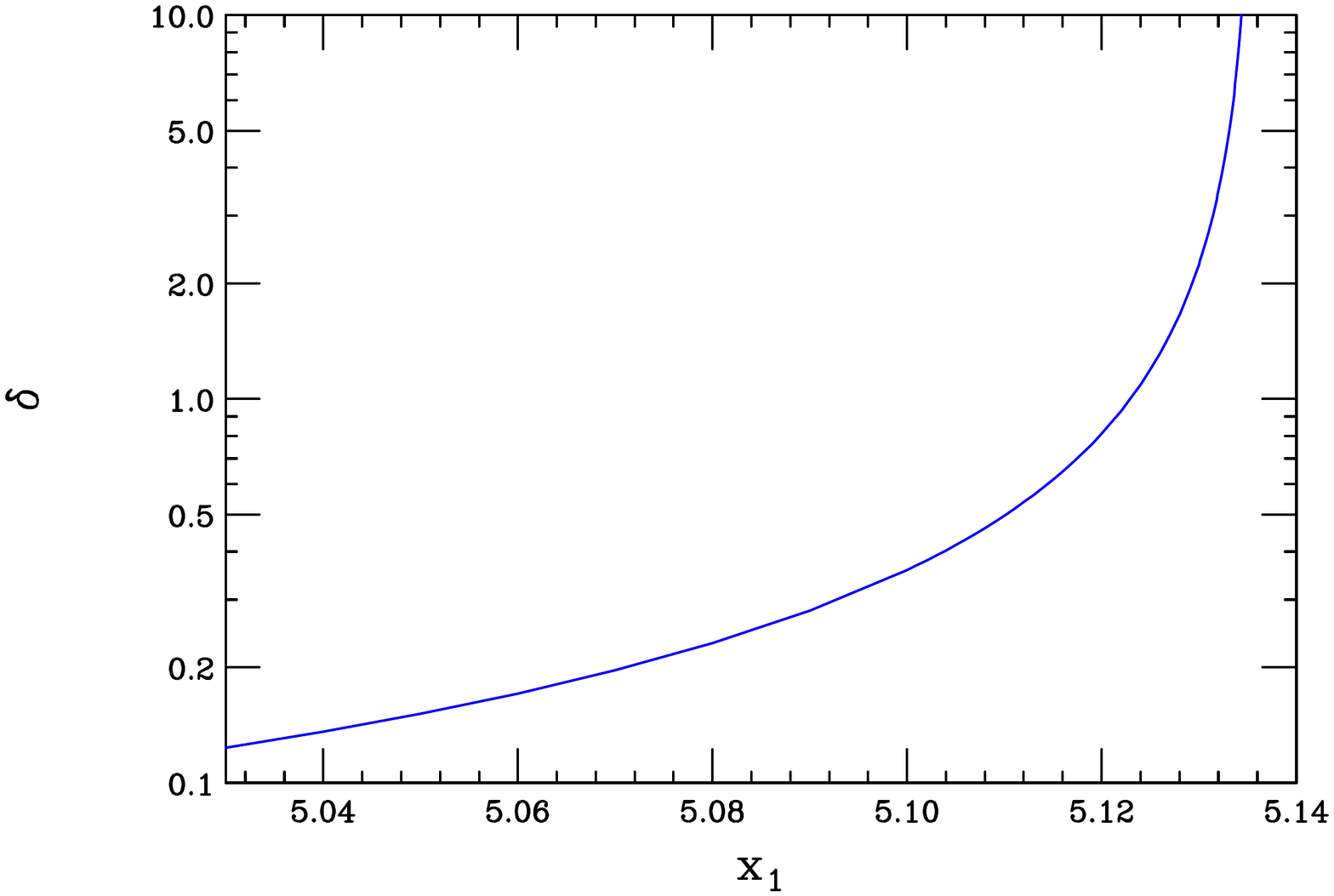}
\includegraphics[scale=0.33,angle=90]{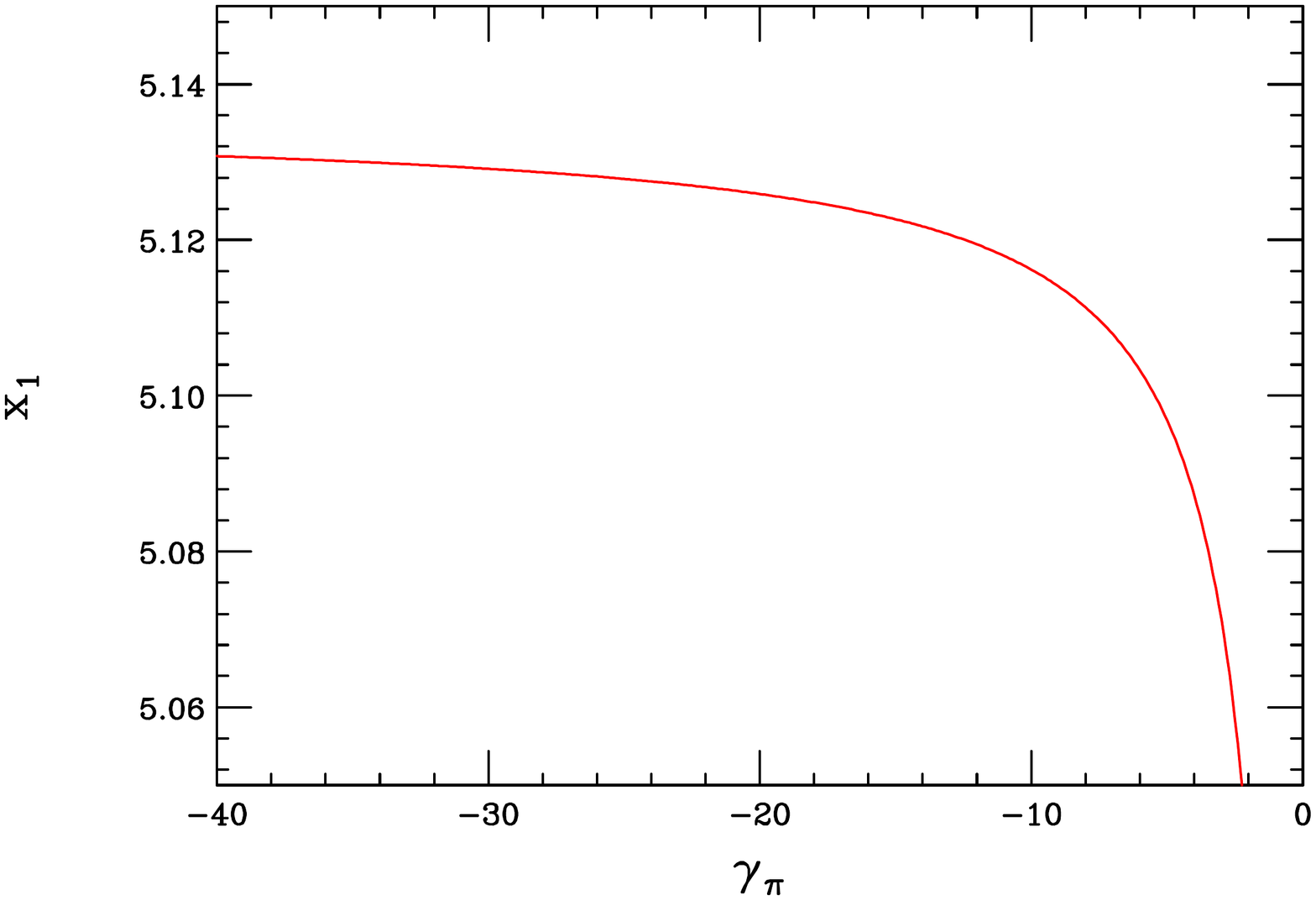}
\vspace*{-1.50cm}
\caption{(Top) Branching fractions for $G_1$ as a function of $\delta$ for, from top to bottom on the left-hand side, top(cyan), 
$W$(blue), $h$(green), $Z$(red), $g$(black) and $\gamma$(magenta) pairs. (Middle) Value of $\delta$ as a function of the first 
graviton root $x^G_1$ in the region of interest. (Bottom) The value of the root $x^G_1$ as a function of $\gamma_\pi$.}
\label{fig1}
\end{figure}

The presence of BLKTs modify the couplings of the graviton KK states to both bulk zero-mode gauge fields and IR-brane  
localized fields.  This leads to an overall suppression of these couplings by a factor of 
\begin{equation}
\lambda_n\equiv \Big[ {{1+2\gamma_0} \over {1+(x^G_n\gamma_\pi)^2-2\gamma_\pi}} \Big]^{1/2}\,.
\end{equation}
In addition, the couplings of the $G_n$ to the transverse zero-mode gauge fields in the bulk are modified by an additional factor 
of \cite{Davoudiasl:2000wi,George}
\begin{equation}
\delta_n = {{2(1-J_0(x^G_n))+(\delta_\pi-\gamma_\pi)(x^G_n)^2 J_2(x^G_n)}\over {(\pi kr_c+\delta_\pi+\delta_0) (x^G_n)^2|J_2(x^G_n))|}} \,.
\end{equation}
We will primarily be interested in $\delta \equiv\delta_1 >0$ in what follows. For the simplicity of argument, we will 
initially ignore the {\it gauge} BLKT contributions to this expression, \ie, we take $\delta_{0,\pi}=0$ for now, and return 
to their influence below. 
The partial widths for the first graviton KK state decays into SM fields, as functions of $\delta$, are now easily 
calculated using the expressions in Ref.~\cite{Giddings:2016sfr} for the IR brane fields (all scaled by overall factors of 
$\lambda_1^2$) and Ref.~\cite{Oliveira:2014kla} for the various gauge field polarizations (which also include overall 
$\lambda_1^2$ factors). The branching fractions, which are independent of the value of $\lambda_1$, are shown in the 
top panel of Fig.~\ref{fig1}. Here we see that these branching 
fractions vary significantly with $\delta$: For small values, decays to IR-brane fields are dominant while decays to the bulk gauge 
fields become more important as $\delta$ increases. A sizable value of B($G_1\to \gamma\gamma)$ is necessary to reproduce the 
observed excess, thus requiring a respectable value for $\delta$.  We note that if $\gamma_\pi=0$, we obtain 
$\delta=(4\pi kr_c)^{-1} \simeq 0.007$ which is far too small.  
Clearly $\delta \simeq 0.5$, or larger, is required (which we take as our benchmark for purposes of demonstration) to obtain an adequate 
branching fraction.  
The graviton branching fractions are tabulated in Table~\ref{gwidths} for this benchmark value. (Fields associated with any  
custodial symmetry are assumed to be too massive to appear in the decays of a 750 GeV state.) 
\begin{table}
\caption{Partial widths and branching fractions for the decay of the first graviton excitation into the SM fields assuming 
$\delta=0.5$} 
\label{gwidths}
\begin{ruledtabular}
\begin{tabular}{c | c | c}
Channel & Scaled partial width~~~ & Branching Fraction~~~ \\ \hline
$\Gamma_{\gamma\gamma}$ &  0.25 $\Gamma_0$ & 4.05\% \\
$\Gamma_{gg}$ &  2.0 $\Gamma_0$  & 32.39\%\\
$\Gamma_{ZZ}$ &  0.37 $\Gamma_0$ & 6.06\%\\
$\Gamma_{WW}$ &  0.73 $\Gamma_0$ & 11.84\%\\
$\Gamma_{hh}$ &  0.12 $\Gamma_0$ & 2.01\%\\
$\Gamma_{b\bar b}$ &  1.5 $\Gamma_0$ & 24.29\%\\
$\Gamma_{t\bar t}$ &  1.2 $\Gamma_0$ & 19.35\% \\
\end{tabular}
\end{ruledtabular}
\end{table}

Noting that $\delta$ is purely a function of $x^G_1$, we next determine the values of $x^G_1$, if any, that yield the 
desired range for $\delta$; this is displayed in the middle panel of Fig.~\ref{fig1}. Here we see that, \eg, 
$x^G_1 \simeq 5.110$ gives $\delta=0.5$. The lower panel in this Figure then shows that this value of $x^G_1$ is realized when the brane term
$\gamma_\pi \simeq -7.652$, thus fixing this quantity for the remainder of this discussion. Employing these results we find 
that $\lambda_1 \simeq 0.025(1+2\gamma_0)^{1/2}$. The partial widths for $G_1$ decays to SM fields now become calculable and 
are also given in Table~\ref{gwidths} in units of 
\begin{equation}
\Gamma_0=\lambda_1^2 {(m^G_1)^3\over {80\pi\Lambda^2_\pi}} = 1.09\times 10^{-3} ~\Biggl[ {{1+2\gamma_0}\over {25}}\Biggr]~
\Biggl[ {5~{\rm TeV}\over\Lambda_\pi} \Biggr]^2 {\rm GeV} \,.
\end{equation}

Armed with the $G_1$ branching fractions for this scenario, we next compute its production cross section at the 
$\sqrt s=8\,, 13$ TeV LHC.  For purposes of demonstration, we will take the 13 TeV diphoton excess cross section to be 
5 fb in our numerical analysis.  Since the graviton couplings to light quarks are effectively absent due to the fermion localizations, 
the graviton is 
produced solely via gluon-gluon fusion.  We find the rate for $G_1$ production and subsequent decay into diphotons to be 
$\sigma_{\gamma\gamma}= 4.86~{\rm fb}~(1+2\gamma_0)/25 ~(5~{\rm TeV}/\Lambda_\pi)^2$ at $\sqrt s = 13$ TeV. Clearly a 
correlated range of $(\gamma_0,\Lambda_\pi)$ near these default values will reproduce the observed cross section. Note that 
taking $\Lambda_\pi=5$ TeV corresponds to a value of $k/{\overline M}_{Pl}=0.029$, similar to those previously considered in 
studies of warped geometries. $\Lambda_\pi \sim 5$ TeV is also remarkably close to the desired value to explain the hierarchy! 
We further note that the $G_1$ state is extremely narrow, \ie, $\Gamma=6.17\Gamma_0$. The jury remains out as to  whether 
the observed excess would best correspond to a wide or narrow resonance.  

\begin{table}
\centering
\caption{Production cross section for 750 GeV $G_1$ for various SM final states at $\sqrt s=8\,, 13$ TeV.} 
\label{grates}
\begin{ruledtabular}
\begin{tabular}{c | c | c}
Channel & $\sigma^{13}$ (fb)~~~ & $\sigma^8$ (fb)~~~ \\ \hline
$\sigma_{\gamma\gamma}$ &  5.0  & 1.18 \\
$\sigma_{gg}$ &  40.0   & 9.44\\
$\sigma_{ZZ}$ &  7.48  & 1.77\\
$\sigma_{WW}$ &  14.6  & 3.45\\
$\sigma_{hh}$ &  2.48  & 0.59\\
$\sigma_{b\bar b}$ &  29.9  & 7.06\\
$\sigma_{t\bar t}$ &  23.4  & 5.64 \\
\end{tabular}
\end{ruledtabular}
\end{table}

\begin{figure}[htbp]
\includegraphics[scale=0.33,angle=90]{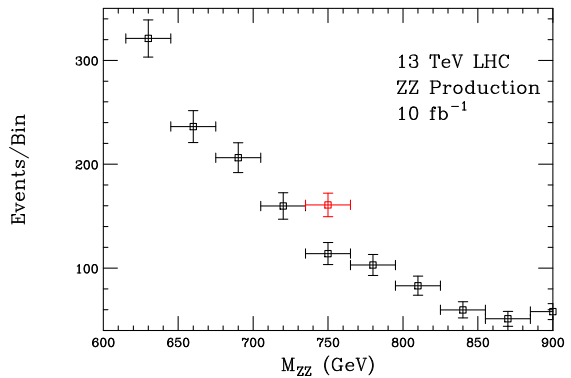}
\vspace*{-1.50cm}
\caption{$ZZ$ invariant mass distribution in the neighborhood of the 750 GeV KK graviton resonance.  The red (black) 
points correspond to the graviton (SM) event rate.
\label{zzzzs}}
\end{figure}

Taking these representative numerics, we show the production rates for a 750 GeV graviton first KK state with subsequent decays 
into SM final states at the $\sqrt s$=8 and 13 TeV LHC in Table~\ref{grates}. We find that the 8 TeV diphoton production 
rate is 1.18 fb, which is consistent with observations by CMS \cite{lhcmoriond}. In particular, the observed ratio of 
cross sections at $M_{\gamma\gamma}=750$ GeV for the CMS spin-2 analysis is $\sigma^{13}/\sigma^{8}=4.2$, agreeing 
with our result. A clear prediction of this scenario  is the existence of of 750 GeV resonance in dijets, $ZZ$, $WW$, 
$hh$, $b\bar b$ and $t\bar t$ final states. The current bounds on a 750 GeV resonance in these various final states are 
collected in Ref.~\cite{Giddings:2016sfr}, and we see from this reference that our predicted rates are well within these 
constraints at both energies.
Clearly it is possible that a 750 GeV $G_1$ resonance may be observed in additional channels with enough luminosity.  
As a simple example, the total event rate for the $ZZ$ final state is shown in Fig.~\ref{zzzzs} for 10 fb$^{-1}$, in comparison 
to the SM background. We see that the graviton signal should be observable above the SM background.  Of course 
incorporating realistic branching fractions, efficiencies and selection cuts would increase the required luminosity 
to observe this peak; this result is meant only to be suggestive and a detailed study 
incorporating these effects should be performed but is beyond the scope of the present work.  However, we note that analyses
from Run I have confirmed that boosted jet techniques can yield high efficiencies in the all hadronic channel, which would boost
statistics.

A necessary signal for a graviton KK tower is the observation of more than one KK state. In our numerical example, the second 
KK graviton, $G_2$, has a predicted mass of $\simeq 1233$ GeV. Unfortunately, this state has a much smaller production cross 
section due to falling parton densities,  a smaller value for $\lambda_2$ and the fact that $\delta_2 \simeq 0.04$ which, 
together imply a significantly smaller $G_2\to gg$ partial width by roughly a factor of $\sim 50$. For this value of $\delta_2$, Fig.~\ref{fig1} further 
implies that this state essentially decays only to IR-brane fields and does not produce a sizable diphoton signal. 
  
\begin{figure}[htbp]
\vspace*{-1.0cm}
\includegraphics[scale=0.33,angle=90]{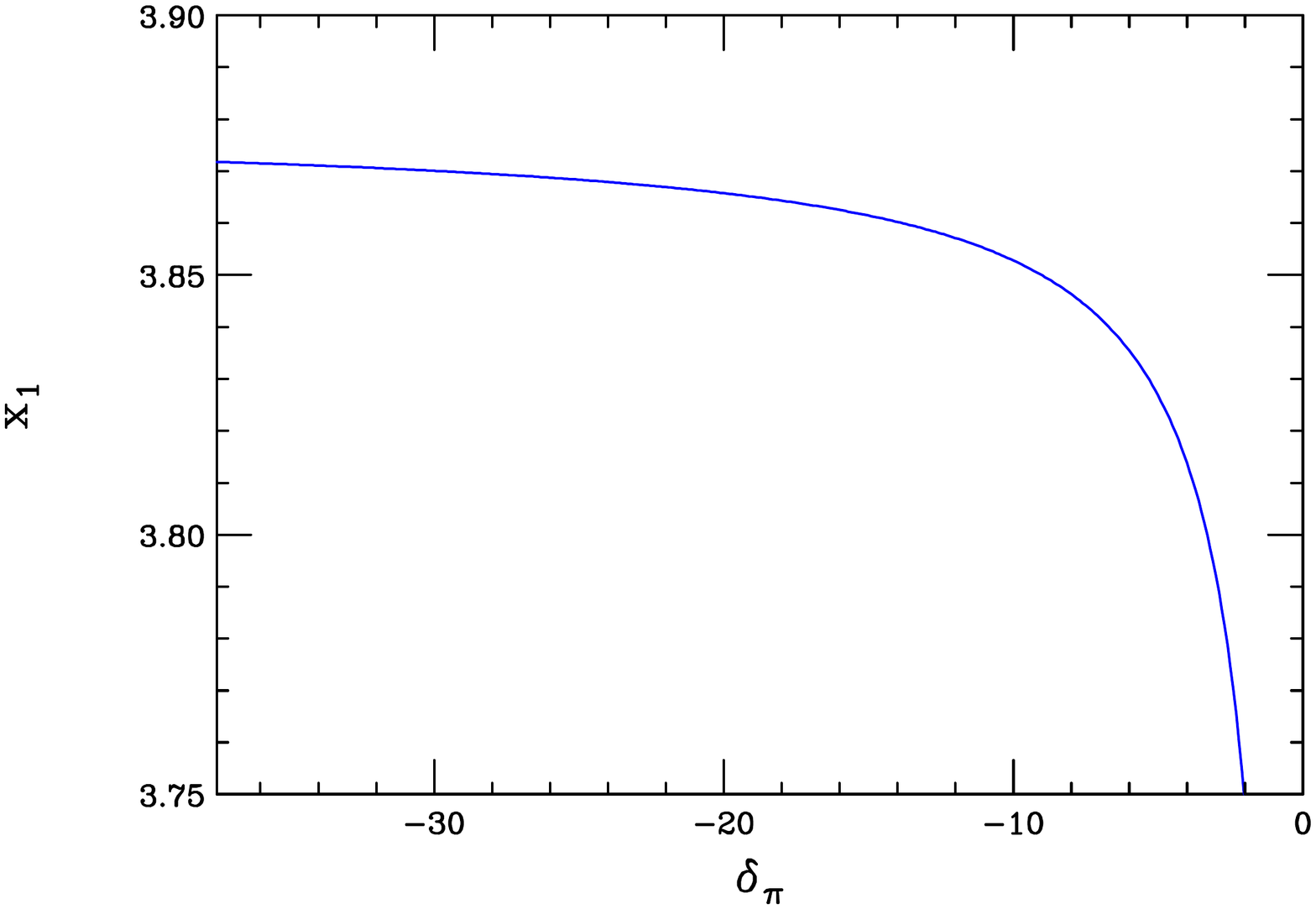}
\includegraphics[scale=0.33,angle=90]{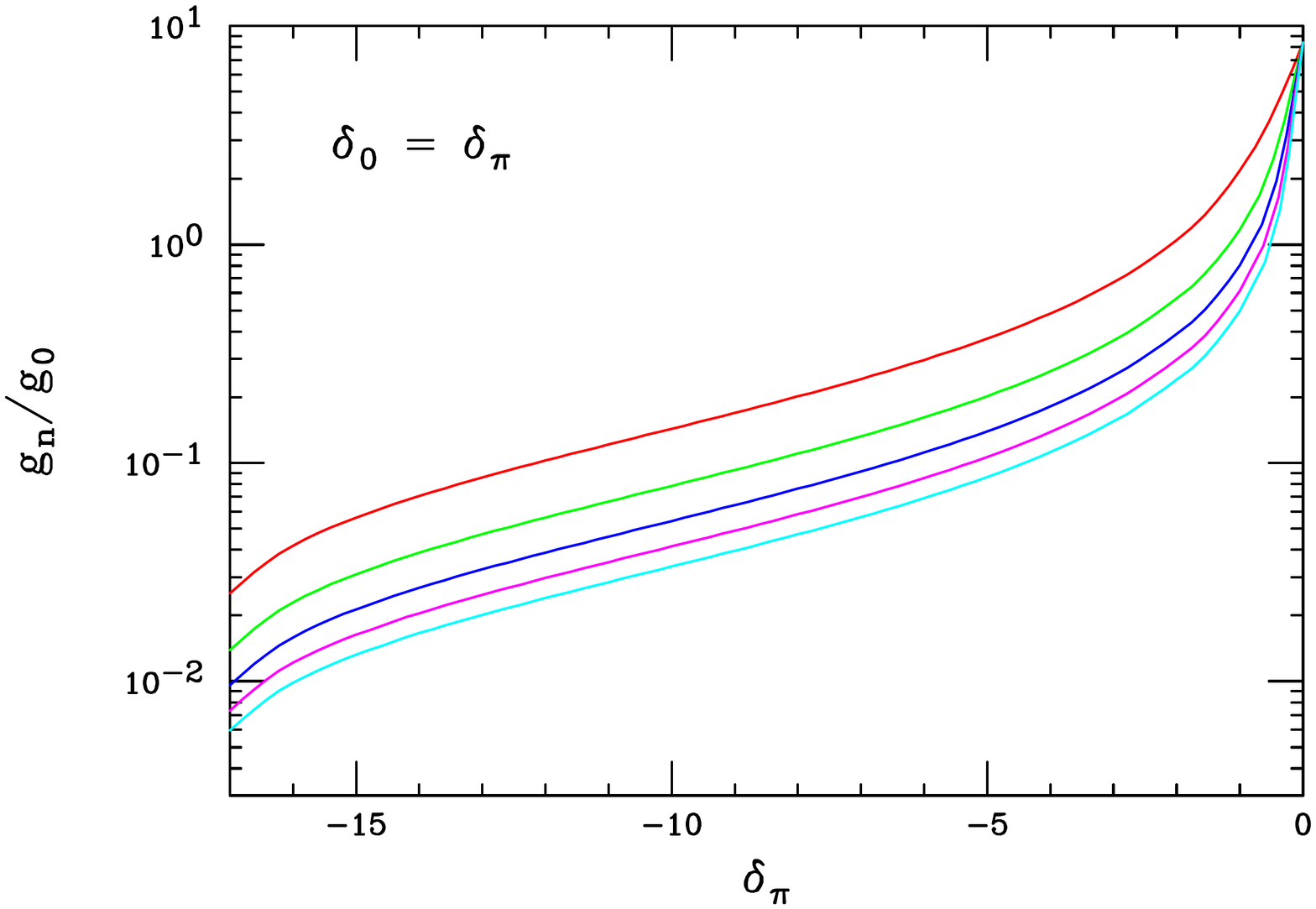}
\includegraphics[scale=0.33,angle=90]{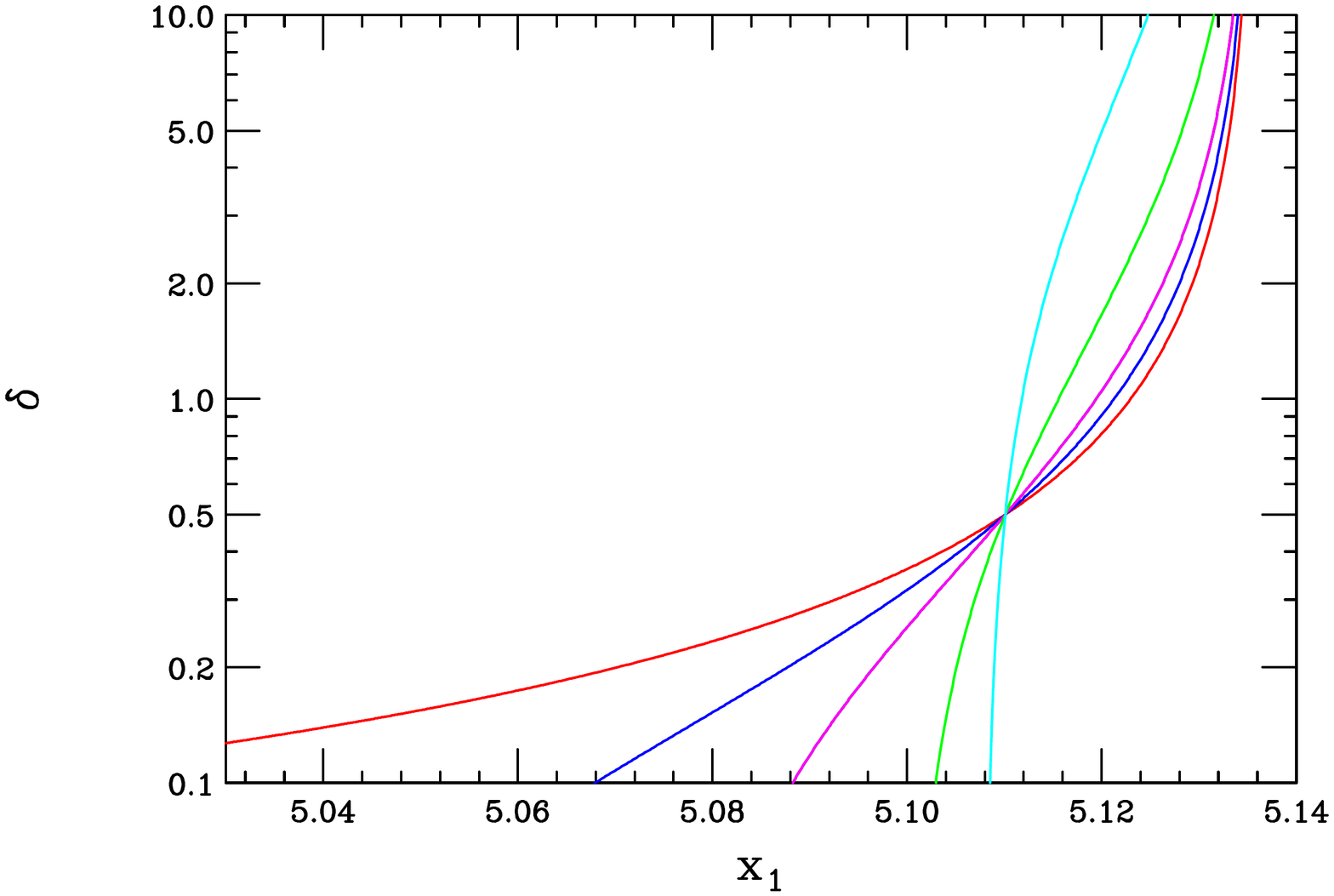}

\vspace*{-1.0cm}
\caption{(Top) Value of $x^A_1$ as a function of $\delta_\pi$. (Middle) Values of the KK gauge couplings scaled to the SM 
value as functions of $\delta_\pi=\delta_0$ for, from top to bottom, n=1-5. (Bottom) The value of $\delta$ as a function of 
$x^G_1$ assuming $\gamma_\pi=-7.65$ for $\delta_\pi=\delta_0$= -5(magenta), -10(blue), -15(green), -17(cyan) or $\gamma_\pi$(red).}
\label{fig3}
\end{figure}

There still remains an issue with this scenario: the lightest set of gauge KK states (collectively called $A_1$ here) have masses below $G_1$, 
even when these fields have their own brane terms\cite{Davoudiasl:2002ua}. As stated above, in the absence of BLKTs the mass of the 
$A_1$ is constrained to lie above $\sim3-4$ TeV from precision electroweak and FCNC constraints, as well as by direct searches. 
However, the existence of BLKTs can substantially reduce the severity of these bounds as they modify the mass spectrum 
and couplings of the gauge KK states, particularly the couplings to IR-brane localized fields.  In our calculations, we 
take the gauge kinetic terms to be equal, \ie, 
$\delta_0=\delta_\pi$, for numerical simplicity.  The 
top panel of Fig~\ref{fig3} shows the root, $x^A_1$, (which governs the mass) of the first gauge KK state   
as a function of $\delta_\pi$,  while the middle panel displays the the scaled couplings of the gauge KK states to the IR-brane fields in SM units. 
Note that the root is independent of $\delta_0$.   A 750 GeV $G_1$ state implies $m^A_{1,(2)} \simeq 565(1033)$ GeV which fixes
 $x^A_1\simeq 3.85$ and $\delta _\pi=-(5-17)$.  It is important to observe from this Figure that the gauge couplings decrease significantly for
 higher gauge KK states. 
 
This scenario avoids potential precision constraints as follows:  ($i$) 
the vev of the Higgs field localized on the IR-brane induces mixing between the gauge fields in the KK tower which subsequently results 
in a shift in the masses of the $W$ and $Z$ bosons, \ie, a non-trivial $\rho$ parameter. This effect on the $W$ and $Z$ masses is 
completely captured by the parameter\cite{Rizzo:1999br} $V\simeq \Sigma_n ~(g_n^2/g_{SM}^2) ~(M_W^2/M_n^2)$, where $g_n,M_n$ are the 
gauge KK couplings to IR-brane fields and the gauge KK masses respectively.  For the range of 
$\delta_\pi$ considered above, especially for the more negative values, it is simple to obtain  values of $V\sim (1-5)\times 10^{-4}$, or even less,
thus suppressing this source of conflict with precision measurements.  The presence of a further custodial symmetry would reduce 
these effects by even larger factors.  ($ii$) A second potential constraint is the introduction of large corrections 
to the $Zb\bar b$ vertex which can arise from two sources: KK excitations of $t-,b-$quarks, which are absent in our scenario, and large KK 
gauge couplings to $t,b$ appearing in loops.  As we can see from the Figure, these couplings are now highly suppressed by the BLKTs.  ($iii$) A 
third issue is FCNC constraints which arise when the gauge KK couplings are both significant and generation dependent. However, if the 
third generation coupling to the lightest gauge KK are suppressed to $\sim 1/10$ their SM value or less {\it and} the first and 
second generation fermions live at $\nu =-c \lsim -0.5$ so that their couplings are similarly suppressed, this issue is ameliorated as all couplings are now both small and roughly equal~\cite{Davoudiasl:2000wi}.  ($iv$) Lastly, the substantial 
suppression of the gauge KK couplings to all of the fermions helps us to avoid any constraints that might arise from 
direct searches, \eg, the potential dilepton signal is reduced by factors of $\sim O(10^3)$. We note that we have not invoked the 
influence of potential fermionic brane terms which can used to further suppress any potentially dangerous fermionic couplings.

A potential concern is that the presence of $\delta_0=\delta_\pi\neq 0$ BLKTs would spoil the graviton results which  
were obtained above in their assumed absence. To address this issue it is sufficient to show that the results obtained 
above are left unmodified, \ie, the values of $\gamma_\pi$ and $x^G_1$ required to obtain $\delta=0.5$ are unaltered 
by the presence of BLKTs in the range of interest. Fixing $\gamma_\pi=-7.652$, we show in the lower panel of Fig.~\ref{fig3} 
the corresponding value of $\delta$ as a function of $x^G_1$ for 4 different values of $-17 \leq \delta_0=\delta_\pi \leq -5$. 
In all cases we find that the required values of $x^G_1 \simeq 5.110$ and $\delta=0.5$ vary by less than $0.01\%$ 
in comparison to the case without BLKTs.  Thus the graviton results 
obtained above are robust for a wide range of interesting values of the gauge BLKTs. Of course solutions can also be 
obtained when the equality of $\delta_0$ and $\delta_\pi$ is no longer assumed.

In conclusion, we examined the scenario of a
warped extra dimension containing bulk SM fields in light of the observed diphoton excess at 750 GeV.  We   
demonstrated that a bulk spin-2 graviton whose action contains localized kinetic brane terms  
is compatible with the excess, while being consistent with all other constraints. BLKTs for the gauge fields 
do not alter these results and, together with appropriate fermion localizations and a possible additional 
custodial symmetry allow these KK states to remain quite light.  The typical value of $\Lambda_\pi$ is found to lie in the region
$\sim 5$ TeV, relevant to the gauge hierarchy.  The 750 GeV signal should eventually be observable in 
other channels and we look forward to future data.

\begin{acknowledgments}
We thank K. Agashe for pointing out an error in the previous version of this manuscript and H. Davoudiasl for discussions.
This work was supported by the Department of Energy, Contract DE-AC02-76SF00515.  In the process of completing this manuscript, a related paper 
appeared\cite{Falkowski:2016glr} which focuses on a somewhat different region of parameter space.
\end{acknowledgments}


\end{document}